\documentclass[iop]{emulateapj}
\usepackage{graphicx,rotating}          
\usepackage{amssymb}
\usepackage{amsmath}
\usepackage{natbib}
\shorttitle{Molecular Tracers of Turbulent Shocks}

\begin{document}

\title{MOLECULAR TRACERS OF TURBULENT SHOCKS IN GIANT MOLECULAR CLOUDS}

\author{A. Pon\altaffilmark{1,2}, D. Johnstone \altaffilmark{2,1}, and M. J. Kaufman \altaffilmark{3,4}}
 
\altaffiltext{1}{Department of Physics and Astronomy, University of Victoria, P.O. Box 3055, STN CSC, Victoria, BC V8W 3P6, Canada; arpon@uvic.ca}
\altaffiltext{2}{NRC-Herzberg Institute of Astrophysics, 5071 West Saanich Road, Victoria, BC V9E 2E7, Canada; Douglas.Johnstone@nrc-cnrc.gc.ca}
\altaffiltext{3}{Department of Physics, San Jose State University, One Washington Square, San Jose, CA 95192-0106, USA; mkaufman@email.sjsu.edu}
\altaffiltext{4}{Space Science and Astrobiology Division, MS 245-3, NASA Ames Research Center, Moffett Field, CA 94035, USA}
 
\begin{abstract}
Giant molecular clouds contain supersonic turbulence and simulations of magnetohydrodynamic turbulence show that these supersonic motions decay in roughly a crossing time, which is less than the estimated lifetimes of molecular clouds. Such a situation requires a significant release of energy. We run models of C-type shocks propagating into gas with densities around 10$^3$ cm$^{-3}$ at velocities of a few km s$^{-1}$, appropriate for the ambient conditions inside of a molecular cloud, to determine which species and transitions dominate the cooling and radiative energy release associated with shock cooling of turbulent molecular clouds. We find that these shocks dissipate their energy primarily through CO rotational transitions and by compressing pre-existing magnetic fields. We present model spectra for these shocks and by combining these models with estimates for the rate of turbulent energy dissipation, we show that shock emission should dominate over emission from unshocked gas for mid to high rotational transitions (J $>5$) of CO.  We also find that the turbulent energy dissipation rate is roughly equivalent to the cosmic-ray heating rate and that the ambipolar diffusion heating rate may be significant, especially in shocked gas.
\end{abstract}


\section{INTRODUCTION}
\label{intro}

	Molecular line observations of giant molecular clouds (GMCs) yield line widths significantly larger than what would be expected from thermal motions alone (e.g., \citealt{Larson81, Solomon87}). These large, nonthermal line widths are generally interpreted as being due to supersonic turbulence, with Mach numbers on the order of 10 (e.g., \citealt{Zuckerman74, McKee07}). Zeeman splitting measurements of magnetic field strengths in molecular clouds show that these supersonic motions are on the order of the Alfv\'{e}n speed, which suggests that magnetohydrodynamic (MHD) waves may play a significant role in molecular clouds \citep{Crutcher99, Crutcher10}. 
		
	Supersonic, hydrodynamic turbulence decays on the order of a free-fall time (e.g., \citealt{Goldreich74, Field78, Elmegreen85}) and thus, maintaining the turbulent support of GMCs for their entire lifetimes, estimated to be between 2 and 30 times longer than the free fall timescale (e.g., \citealt{Mouschovias76, Shu77, Blitz80, Shu87, Williams97, Elmegreen00, Hartmann01, MacLow04}), is a significant problem. Based on theoretical calculations \citep{Arons75}, it was believed that MHD turbulence would decay an order of magnitude slower than hydrodynamic turbulence, thereby preventing the dissipation of turbulent energy in GMCs; however, simulations of MHD turbulence show that MHD turbulence also decays on the order of a free-fall time at the driving scale \citep{Gammie96, MacLow98, Stone98, MacLow99, Padoan99, Ostriker01}. 
		
	In MHD turbulence simulations, turbulent energy is dissipated via numerical viscosity and artificial viscosity in shock fronts. Under the assumption that the dissipated turbulent energy is lost as heat and rapidly radiated away, many MHD simulations are run with isothermal equations of state and thus, these simulations do not explicitly follow where the dissipated turbulent energy goes (e.g., \citealt{Stone98, SmithZuev00}). \citet{Basu01} made a first attempt to compare the CO $\mbox{J} = 1 \rightarrow 0$ luminosities of molecular clouds to what they predicted would be seen from molecular clouds based upon simple energetic arguments. Since then, however, little progress has been made in determining where this turbulent energy goes and whether there are any observational signatures of this dissipated energy.
	
		Shocks increase the temperature and density of the shocked gas, which, in turn, can substantially alter the chemistry of the gas and the emission coming from the gas (e.g., \citealt{Kaufman96I, Kaufman96II}), thereby potentially providing a distinct signature and tracer of turbulent energy dissipation via shocks. For typical turbulent velocities and magnetic field strengths of molecular clouds, the magnetic field is capable of transmitting information about the presence of a shock to ions upstream of the shock front. This eliminates discontinuities in gas properties across the shock front and spreads out the thickness of the shock. In turn, this leads to lower temperatures in the shocked gas and prevents molecules from being dissociated. Such a shock is referred to as a continuous, or C-type, shock and is described in more detail in \citet{Mullan71}, \citet{Draine80}, and \cite{Draine93}. 
		
	We run models of C-type, MHD shocks, based upon \citet{Kaufman96II}, propagating into molecular gas with densities around 10$^3$ cm$^{-3}$ at velocities of a few km s$^{-1}$, appropriate for the ambient conditions inside of a molecular cloud, to determine which species and transitions dominate the cooling and radiative energy release associated with shock cooling in turbulent molecular clouds. The shock velocities modeled are on the order of the typical turbulent velocity of molecular clouds, which are much lower than the velocities of protostellar outflows that have been the target of previous studies (e.g., \citealt{Chernoff82, Timmermann96, Kaufman96I, Kaufman96II}). These shock models are combined with estimates for the rate of turbulent energy dissipation in molecular clouds to predict the integrated intensities of various shock excited lines coming from an entire molecular cloud and these integrated intensities are then compared to those from photodissociation region (PDR) models based upon \citet{Kaufman99}. 
	
	Typical scaling relations of GMCs are presented in Section \ref{scaling} and the turbulent energy dissipation rate of molecular clouds is derived in Section \ref{lturb}. The shock and PDR models used in this paper are described in Sections \ref{code}-\ref{pdr} and the results of these models are presented in Section \ref{results}. In Section \ref{discussion}, the implications of these results are discussed, and in Section \ref{heating}, the rate of turbulent dissipation is compared to other known heating mechanisms in molecular clouds. Finally, our findings are summarized in Section \ref{conclusions}.
		
\section{SETUP}
\label{setup}

\subsection{Scaling Relations of GMCs}
\label{scaling}

Correlations between the size, density, and line-of-sight velocity dispersion of GMCs, as determined through CO observations, are well known and are collectively referred to as Larson's laws (e.g., \citealt{Larson81,Solomon87, Heyer04}). The best fitting scaling relations found by \citet{Solomon87} are:
\begin{eqnarray}
\label{eqn:larson1}
\sigma&=&0.72 \left(R/\mbox{pc}\right)^{0.5} \mbox{ km s}^{-1},\\
\label{eqn:larson2}
\rho&=&134\left(R/\mbox{pc}\right)^{-1} \mbox{M$_\odot$ pc}^{-3},
\end{eqnarray}
where R is the effective radius (the radius of a  spherical cloud with the same projected surface area as the observed cloud), $\sigma$ is the one-dimensional velocity dispersion (which we assume is equal to the observed line-of-sight velocity dispersion), and $\rho$ is the average density. These relations suggest that a cloud with a mean density of 10$^3$ cm$^{-3}$ has a radius of approximately 2 pc, a mass of 2000 M$_\odot$, a total molecular hydrogen column density of $1.2 \times 10^{22}$ cm$^{-2}$ (corresponding to a visual extinction of 12 through the entire cloud), and a one-dimensional velocity dispersion of about 1 km s$^{-1}$.

The size-velocity relationship is fairly well established, although there is some evidence that the velocity dispersion of a molecular cloud may also depend upon the column density of that cloud \citep{Heyer09}. The validity of the size-density relation, however, is much less certain, as the observed relationship may be only due to the limited dynamical range of current observations \citep{BallesterosParedes02}. For the shock models used in this paper, Larson's laws are only used to confirm that the simulated parameter range roughly corresponds to the properties of observed molecular clouds. Neither part of Larson's laws is used to calculate the integrated intensity of the shock emission.

	\citet{Solomon87} also found a correlation between the velocity dispersion and $^{12}$CO J = 1 $\rightarrow$ 0 total luminosity, L$_{1 \rightarrow 0}$, of molecular clouds. In units of K km s$^{-1}$ pc$^{-2}$, \citet{Solomon87} found that the CO 1 $\rightarrow$ 0 luminosity is:
\begin{equation}
\label{eqn:Lsol}
L_{1 \rightarrow 0} = 130 \sigma^5.
\end{equation}

The relationship between the magnetic field strength in a molecular cloud, B, and the number density of hydrogen nuclei, $n_H$, is often expressed in the form
\begin{equation}
B = b\,  n_H^{k} \, \mu \mbox{G},
\label{eqn:bfield}
\end{equation}
where b and k are the fitting parameters. A value of k = 0.5 corresponds to a constant magnetic energy density \citep{McKee07} and is expected from ambipolar diffusion collapse models \citep{Fiedler93}. A k = 0.5 relation is also expected if the turbulent velocity in a cloud is always roughly the Alfv\'{e}n speed (e.g., \citealt{Crutcher99}). A value of k = 2/3, however, is predicted if magnetic fields are unimportant and a molecular cloud is able to maintain a roughly spherical shape during its collapse \citep{Mestel66, Crutcher99, Crutcher10}. 

\citet{Crutcher99} compiled Zeeman splitting observations and found b = 0.95 and k = 0.5 \citep{McKee07}. The MHD simulations of \citet{Padoan99} exhibit a k = 0.4 relation and the relation $\mbox{b} = 1$, $\mbox{k} = 0.5$ is commonly adopted (e.g., \citealt{Draine83, Kaufman96I,Kaufman96II}). Recently, \citet{Crutcher10} have examined all of the available Zeeman splitting observations, including those used by \citet{Crutcher99}, and found that the best fit for the maximum observed line-of-sight magnetic field strength comes from the relation:
\begin{equation}
B_{max} = 
\begin{cases} 
10 \, \mu G & n_H < 300 \mbox{ cm}^{-3}\\[0.15 in]
10 \mu G \left(\frac{n_H}{300 \, cm^{-3}}\right)^{0.65} & n_H \ge 300 \mbox{ cm}^{-3}.
\end{cases}
\end{equation}
For densities greater than 300 cm$^{-3}$, the above relation corresponds to k = 2/3 and b = 0.25. \citet{Crutcher10} also note that their data are consistent with having line-of-sight magnetic field strengths down to essentially zero.

The maximum magnetic field strengths that were fit by \citet{Crutcher10} most likely correspond to cases where the magnetic field is highly aligned with the line-of-sight, such that the full magnetic field strength is measured. The average magnetic field strength along any random direction will thus be only half of the strength given by the above relation. For a cloud with an H$_2$ density of 10$^3$ cm$^{-3}$, the \citet{Crutcher10} relation therefore predicts that the average magnetic field strength along any random direction is 17 $\mu$G. Intrinsic scatter in the magnetic field strength between different clouds will also likely further reduce the average magnetic field strength. 

\subsection{Turbulent Energy Dissipation Rate}
\label{lturb}

The turbulent energy density of a molecular cloud is approximately
\begin{equation}
E_{turb} = \frac{3}{2} \, \rho \sigma^2. 
\end{equation}
Following the discussion in \citet{Basu01}, the mean turbulent energy dissipation rate per volume can be written as $\Gamma_{turb}=E_{turb}/t_d$, where $t_d$ is the dissipation timescale. We define the flow crossing time of the cloud as $t_c=2R / \sigma$ and introduce the ratio of the dissipation time to the flow crossing time as a new parameter: $\kappa = t_d / t_c$. The turbulent dissipation rate per volume is thus
\begin{equation}
\label{eqn:gammaturb2}
\Gamma_{turb}=\frac{3 \, \rho \, \sigma^3}{4 \, \kappa \, R}.
\end{equation}

As shown in \citet{Basu01}, rather than writing the turbulent energy dissipation rate in terms of $\kappa$, the dissipation rate can be expressed in terms of the driving scale of the turbulence, $\lambda$:
\begin{equation}
\Gamma_{turb}=\eta \frac{\rho \, \sigma^3}{\lambda},
\label{eqn:gammaeta}
\end{equation}
where $\eta$ is a dimensionless parameter that is a function of the density, velocity dispersion, and driving wavelength. Comparing Equation (\ref{eqn:gammaturb2}) to Equation (\ref{eqn:gammaeta}) gives the relation
\begin{equation}
\kappa = \frac{3 \, \lambda}{4 \, \eta \, R}.
\label{eqn:lambdatokappa}
\end{equation}

Periodic box simulations of MHD turbulence have found that for a variety of initial conditions, $\eta$ has a value between 0.5 and 4 \citep{Gammie96, Stone98, MacLow99, Ostriker01}. Unfortunately, there is no clear consensus on what scale turbulence is driven on. Protostellar outflows, which drive turbulence on small scales, appear to have enough energy to drive turbulence in active star forming regions \citep{Quillen05, Curtis10, Arce10}. It is, however, unclear whether outflows are capable of driving turbulence across an entire molecular complex \citep{Banerjee07, Arce10}. Studies of density and velocity structure in molecular clouds find that the observed structures are only consistent with driving at size scales at, or above, the size of the cloud (e.g., \citealt{Ossenkopf02, Brunt03, Heyer04, Brunt09, Padoan09}). Supersonic turbulence has also been observed in the Polaris Flare, which is devoid of any protostars \citep{Andre10}. 

For the remainder of this paper, a $\kappa$ value of one will be adopted, for which the turbulent dissipation timescale is equal to the flow crossing time of the cloud. Via Equation (\ref{eqn:lambdatokappa}) and the numerical factors for $\eta$, this corresponds to a turbulent driving scale on the order of the size of a molecular cloud. 

Equation (\ref{eqn:gammaturb2}) is a general result that can be applied to any cloud, given that the characteristic radius, density, and velocity dispersion are known. For this paper, the simplifying assumption that clouds are spherical will be made, such that the total turbulent energy dissipation rate is 
\begin{eqnarray}
L_{turb} &=& \frac{3 \, \rho \, \sigma^3}{4 \, \kappa R} \frac{4 \pi \, R^3}{3},\\
\label{eqn:finallturb}
L_{turb} &=& \frac{\pi \, \rho \, \sigma^3 \, R^2}{\kappa}.
\end{eqnarray}

If all of the dissipated turbulent energy is radiated away, the corresponding total integrated intensity, $I_{turb}$, is 
\begin{eqnarray}
I_{turb} &=& \frac{L_{turb}}{4 \pi^2 \, R^2},\\
I_{turb} &=& \frac{\rho \, \sigma^3}{4 \pi \, \kappa}.
\end{eqnarray}
This integrated intensity is independent of the size of the molecular cloud. For this paper, a mean mass per particle of $4.6 \times 10^{-24}$ g, or about 2.77 amu, is adopted, and with this value, the above equation becomes
\begin{eqnarray}
I_{turb} &=& 3.66 \times 10^{-7} \, \kappa^{-1} \left(\frac{n}{10^3 \mbox{ cm}^{-3}}\right) \left(\frac{\sigma}{1 \mbox{ km s}^{-1}}\right)^3 \nonumber\\
              && \mbox{ erg s}^{-1} \mbox{ cm}^{-2} \mbox{ steradian}^{-1}, \\
\label{eqn:intturb}
\nonumber \\
I_{turb} &=& 8.60 \times 10^{-18} \, \kappa^{-1} \left(\frac{n}{10^3 \mbox{ cm}^{-3}}\right) \left(\frac{\sigma}{1 \mbox{ km s}^{-1}}\right)^3 \nonumber \\
              && \mbox{ erg s}^{-1} \mbox{ cm}^{-2} \mbox{ arcsec}^2.
\end{eqnarray}

\subsection{Shock Code}
\label{code}

	To determine which species and transitions dominate the cooling and radiative energy release associated with shock cooling of turbulent molecular clouds, we run models of C-type shocks based upon the models of \citet{Kaufman96II} with initial conditions corresponding to that expected for roughly one parsec sized molecular clouds.  

	The code used first calculates the temperature, density, chemical abundance, and velocity profiles of a C-type shock. To do so, it calculates the cooling rates for rotational and vibrational transitions of H$_2$O, H$_2$, and CO \citep{Neufeld93}; collisions between the neutral gas and cooler dust grains \citep{Hollenbach89}; and H$_2$ dissociative cooling \citep{Lepp83, Hollenbach89}. The freezing out of molecules on dust grains is, however, not calculated by the code as the freezeout timescale is expected to be much longer than the shock cooling timescale. 
		
	Once the shock structure is determined, the code then calculates the integrated intensities of each molecular transition of interest by solving the partial differential equations for the line emission at each point and then integrating the emission over the entire shock profile.  This extra step of determining individual line strengths, rather than just determining an overall cooling rate for a particular molecule, is only used for CO.  In this later step, the code only includes rotational line emission for gas down to 10 K, unlike in \citet{Kaufman96II}, where emission is only included from gas above 50K. No such temperature limitation is used when calculating the overall cooling rates for each molecule in the first half of the code. For a more detailed description of how this code works, please see \citet{Kaufman96II}.	
			
	By equating the total kinetic energy dissipated by shocks to the turbulent energy dissipation rate of a molecular cloud, given by Equation (\ref{eqn:finallturb}), we scale our shock models to predict the expected integrated intensities from each CO rotational line. That is, the integrated intensity of each line is set to the appropriate fraction of the integrated intensity given by Equation (\ref{eqn:intturb}). It is assumed that the shock emission is coming from a region larger than the size of the beam and each line is scaled equally under the assumption that all of the lines are optically thin. The effect of the lower lying lines being optically thick is discussed further in Section \ref{caveats}. 
		
\subsection{Shock Code Parameters}
\label{parameters}
		
	For each shock model, the same, roughly solar, chemical composition as used in \citet{Kaufman96I, Kaufman96II} is used. In particular, the initial CO number density is set to be $1.2 \times 10^{-4}$ times that of the H nuclei number density and the initial H$_2$O abundance is set to 10$^{-7}$. As shown in Section \ref{scaling}, a one parsec sized molecular cloud is expected to have a density of approximately 10$^3$ cm$^{3}$. Thus, the initial H$_2$ density is set to be either 10$^{2.5}$, 10$^3$, or 10$^{3.5}$ cm$^{-3}$ in these models.

If the velocity distribution of gas particles in a molecular cloud is Gaussian in every direction, with a one-dimensional velocity dispersion of $\sigma$, then the distribution of relative velocities between two gas particles in the cloud will also be Gaussian in every direction with a one-dimensional velocity dispersion of $\sqrt{2} \sigma$. Since the energy dissipation rate of a shock scales with the third power of the shock speed, the mean speed at which energy is dissipated is the cube root of the mean cubed velocity difference between two gas particles, $<\Delta v^3>^{1/3}$, which is roughly 2.4$\sigma$. The shock velocity at which the peak energy dissipation rate occurs is slightly higher, approximately 3.2$\sigma$. Thus, the characteristic shock velocity in a molecular cloud with a one-dimensional velocity dispersion of 1 km s$^{-1}$, consistent with the size-velocity relation for a radius of 1 pc, is on the order of 2-3 km s$^{-1}$. For the remainder of this paper, we assume that the one-dimensional velocity dispersion is a factor of 3.2 smaller than the shock velocity. The larger conversion factor of the two mentioned above is chosen so that the corresponding velocity dispersions, and thus the shock-integrated intensities calculated in Section \ref{model emission}, are smaller.

Models with shock velocities of 2 and 3 km s$^{-1}$ are computed. For a temperature of 10 K, these velocities correspond to Mach numbers of 12 and 17, respectively. While these velocities are appropriate for turbulent motions in a molecular cloud, they are much lower than the typical velocities of protostellar outflows and winds. Such higher velocity flows have been modeled extensively in the past and give rise to significantly higher post shock temperatures (e.g., \citealt{Kaufman96I, Kaufman96II}).

The strength of the magnetic field parallel to the shock front is initialized using the parameterizations k = 0.5 and b = 0.1 or 0.3, where b and k are as defined in Equation (\ref{eqn:bfield}). The component of the magnetic field perpendicular to the shock front is always set to zero, as this component has no effect on the shock structure in our steady state, plane parallel models. Thus, the initial magnetic field strength ranges from 3 $\mu G$ to 24 $\mu G$ in the different models. For a weaker field parallel to the shock front, the shock thickness is smaller and the energy released in line radiation is relatively larger. This is why magnetic field strengths that are slightly lower than, although still generally consistent with, the average line-of-sight magnetic field strength given by the scaling relation of \citet{Crutcher10} have been chosen. 

The Alfv\'{e}n speed is 
\begin{equation}
v_{A} = \frac{B}{\sqrt{4 \pi \, \rho}}.
\end{equation}
For our shock models, the Alfv\'{e}nic Mach number, given by $M_A = v_{shock}/v_A$, ranges from 4 to 16. 

Twelve shock models are run, one for each combination of initial density, shock velocity, and magnetic field b parameter. Table \ref{table:initial} gives the shock velocity, magnetic field strength, Mach number, and Alfv\'{e}nic Mach number for each model. A naming convention of nWXvYbZ is adopted, where W.X is the logarithm of the initial H$_2$ number density in cm$^{-3}$, Y is the shock velocity in km s$^{-1}$, and Z is the magnetic b parameter. 

\begin{deluxetable}{ccccccc}
\tabletypesize{\scriptsize}
\tablecolumns{7}
\tablecaption{Shock Model Properties \label{table:initial}}
\tablewidth{0pt}
\tablehead{
\colhead{Model} & \colhead{log(n)} & \colhead{v} & \colhead{b} & \colhead{B} & \colhead{Mach} & \colhead{$M_A$} \\
\colhead{} & \colhead{(cm$^{-3}$)} & \colhead{(km s$^{-1}$)} & \colhead{} & \colhead{($\mu$G)} & \colhead{} & \colhead{} \\
\colhead{(1)} & \colhead{(2)} & \colhead{(3)} & \colhead{(4)} & \colhead{(5)} & \colhead{(6)} & \colhead{(7)}
}
\startdata
n25v2b1 & 2.5 & 2 & 0.1 & 3 & 12 & 11 \\
n25v3b1 & 2.5 & 3 & 0.1 & 3 & 17 & 16 \\
n25v2b3 & 2.5 & 2 & 0.3 & 8 & 12 & 4 \\
n25v3b3 & 2.5 & 3 & 0.3 & 8 & 17 & 5 \\
n30v2b1 & 3 & 2 & 0.1 & 4 & 12 & 11 \\
n30v3b1 & 3 & 3 & 0.1 & 4 & 17 & 16 \\
n30v2b3 & 3 & 2 & 0.3 & 13 & 12 & 4 \\
n30v3b3 & 3 & 3 & 0.3 & 13 & 17 & 5 \\
n35v2b1 & 3.5 & 2 & 0.1 & 8 & 12 & 11 \\
n35v3b1 & 3.5 & 3 & 0.1 & 8 & 17 & 16 \\
n35v2b3 & 3.5 & 2 & 0.3 & 24 & 12 & 4 \\
n35v3b3 & 3.5 & 3 & 0.3 & 24 & 17 & 5 
\enddata
\tablecomments{Column 1 shows the model name, while Columns 2 and 3 show the logarithm of the initial density and shock velocity of each model respectively. Column 4 represents the magnetic b parameter, as defined in Equation (\ref{eqn:bfield}), and Column 5 shows the resulting initial magnetic field strength. Columns 6 and 7 show the Mach number and Alfv\'{e}nic Mach number of the models, respectively.}
\end{deluxetable}

\subsection{PDR Model}
\label{pdr}
The shocked gas in a molecular cloud is not the only source of molecular line emission. The cool, well-shielded gas and the warmer gas in the PDR at the cloud's surface, which is exposed to the interstellar radiation field (ISRF), will also contribute emission. To model this emission from unshocked gas, PDR models based on \citet{Kaufman99} are used. As suggested by \citet{Kaufman99}, these plane parallel models are adjusted for spherical geometry by using the equation
\begin{equation}
L = \int 4 \pi j(N) r^2 dr,
\end{equation}
where r is the radial distance from the center of the cloud and j(N) is the emissivity at a column N from the surface of the cloud. \citet{Kaufman99} estimate that this procedure produces results that are within a factor of 1.5 from intrinsically spherical PDR models. Furthermore, for any optically thin line, the resulting integrated intensity is doubled to account for photons originally emitted radially inward.

The PDR models used have ISRFs of 3 Habing, where the average far ultraviolet ISRF in free space is 1.7 Habing or $2.7 \times 10^{-3}$ erg cm$^{-2}$ s$^{-1}$ \citep{Tielens05}, and microturbulent Doppler line widths of 1.5 km s$^{-1}$, similar to the velocity dispersions of the shock models. It is assumed that the PDR emission fills the beam. These PDR models do not take into account the freezing out of CO onto dust grains.

A density of 10$^3$ H$_{nuclei}$ cm$^{-3}$ is used for all of the comparison PDR models, which is comparable to the median initial density in the shock models. We believe that this is an appropriate comparison density for the 10$^{3.5}$ cm$^{-3}$ shock models because a density gradient should be present within realistic molecular clouds, with the density decreasing toward the periphery of the cloud, in order for the clouds to remain in pressure equilibrium. Thus, it is expected that the warm outer layers of molecular clouds, from which most of the PDR emission comes from, are at lower densities than the bulk of the cool, CO-rich gas in the interior of the clouds from which most of the shock emission originates. In PDR models with densities below 10$^3$ H$_{nuclei}$ cm$^{-3}$, CO only forms in the gas phase once the gas has cooled significantly, such that there is almost no emission in the mid to high J lines of CO. To be more conservative in our findings of when shock emission is stronger than PDR emission, we prefer using PDR models with densities of 10$^3$ H$_{nuclei}$ cm$^{-3}$  for comparison with the 10$^3$ and 10$^{2.5}$ cm$^{-3}$ shock models, even though these PDR models may over predict the CO emission from 10$^{2.5}$ cm$^{-3}$ gas. The results should be roughly consistent for the 10$^3$ cm$^{-3}$ shock models.

For the comparison PDR models, the size of the molecular cloud must also be known in order to know at what A$_V$ to cut the model. For each shock model, the \citet{Solomon87} size-velocity relation is used to determine an appropriate, typical size for a cloud and then the CO column density of that cloud is determined under the assumptions that the cloud is spherical and has a typical CO abundance of $2 \times 10^{-4}$ \citep{Glover10} throughout the entire cloud. This CO column density is then used to determine the appropriate depth of the comparison PDR model such that the PDR model has the same CO column density. The depths of the PDR models are chosen based upon a CO column density, rather than upon a hydrogen nuclei column density, because the initial chemical abundances used for the shock models are consistent with gas in which CO has already formed in the gas phase and because CO cooling dominates the energy budget of the shock models, as described later in Section \ref{model emission}. 

Two shock models, n35v3b1 and n35v3b3, require CO column densities larger than the total CO column present at the maximum depth of the 10$^3$ cm$^{-3}$ PDR model. The contribution from the most deeply embedded layers of this PDR model to the total emergent flux does, however, drop to negligible values for all of the CO lines. This is because the lower lines are optically thick and the gas temperature at high A$_V$ is too low for any significant emission in the higher lines. Thus, we use the full extent of the 10$^{3}$ cm$^{-3}$ PDR model as the comparison model for these two shock models and we do not believe that this failure to exactly match the CO column densities of these two models significantly affects our results.

\subsection{Empirical CO 1 $\rightarrow$ 0 Luminosities}
\label{solomon}
	As described in Section \ref{scaling}, \citet{Solomon87} found an empirical correlation between the velocity dispersion and $^{12}$CO J = 1 $\rightarrow$ 0 total luminosity of molecular clouds. For the velocity dispersion corresponding to each of the shock models, the 1 $\rightarrow$ 0 integrated intensity expected from this \citet{Solomon87} relation is calculated using a cloud radius from the \citet{Solomon87} size-velocity relation and the assumption that molecular clouds are spherical. While we have used the size-velocity relationship in determining the comparison PDR spectra and in calculating the empirically expected 1 $\rightarrow$ 0 integrated intensity, we re-emphasize that this relationship is not used at any time in the calculation of the expected shock spectra.	

\section{RESULTS}
\label{results}

\subsection{Shock Line Emission}
\label{model emission}

\begin{figure}[htbp] 
   \centering
   \includegraphics[width=3.5 in]{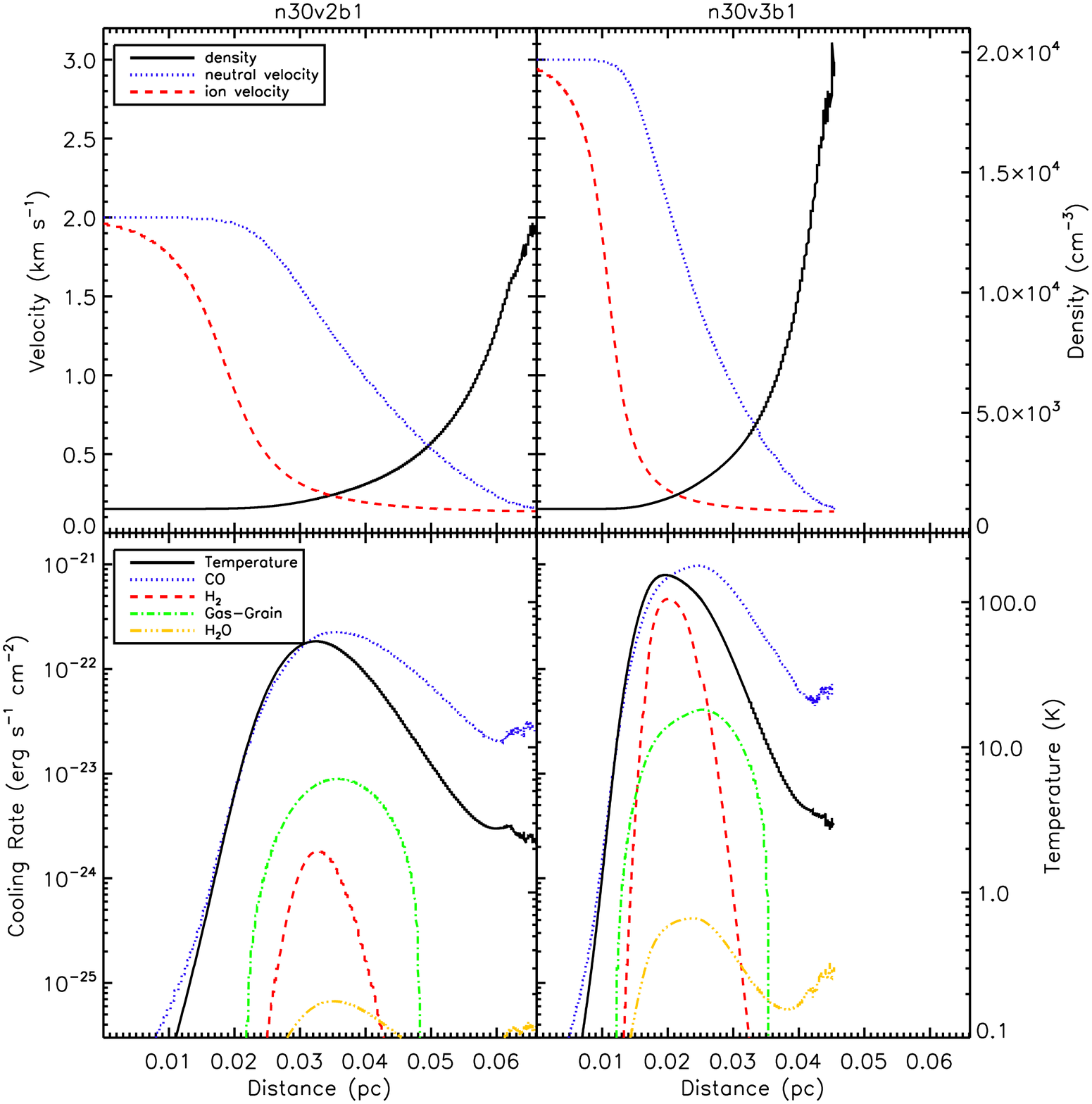}
   \caption{Various profiles of models n30v2b1 and n30v3b1. The top row shows density, neutral velocity, and ion velocity profiles as the solid (black), dotted (blue), and dashed (red) lines, respectively. The velocity axis is given on the left-hand border while the density axis is given on the right-hand border. The bottom row shows temperature profiles as the solid (black) lines and cooling profiles due to CO, H$_2$, gas-grain interactions, and H$_2$O as the dotted (blue), dashed (red), dash-dotted (green), and dash-triple-dotted (yellow) lines, respectively. The cooling rate axis is given on the left border and the temperature axis is given on the right border. The CO cooling profiles have been boxcar smoothed beyond a distance of 0.06 pc in model n30v2b1 and past  0.04 pc in model n30v3b1 due to the presence of high frequency noise. This noise is likely numerical in nature and should not significantly affect our results (see the text in Section \ref{model emission}). The left-hand column shows profiles of the n30v2b1 model and the right-hand column shows profiles of the n30v3b1 model. The x-axes of all four boxes are the same and the y-axis scaling is the same for both models. Please see the online version for a color version of this figure.}
   \label{fig:profiles}
\end{figure}

	In the upper panels of Figure \ref{fig:profiles}, the neutral velocity, ion velocity, and density profiles of models n30v2b1 and n30v3b1 are shown. As typical in MHD, C-type shocks, the density and velocity profiles show no sharp discontinuities and the ion velocity decreases before the neutral velocity does. Due to mass conservation, the density and neutral velocity are inversely related to each other, and thus, the maximum density reached in model n30v3b1 is larger than that reached in model n30v2b1. The magnetic field strength and ion velocity are similarly inversely correlated. 

	The lower panels of Figure \ref{fig:profiles} show the temperature profiles of models n30v2b1 and n30v3b1 as well as the cooling profiles due to CO, H$_2$, and H$_2$O lines and gas-grain coupling. During the initial stages of a shock, where the gas temperature is still increasing, the rate of CO cooling increases in close tandem to the increase in temperature. After the gas has reached its peak temperature, the CO cooling rate remains higher than it was at the same temperature earlier in the shock. This is due to the higher gas densities in these later regions of the shock that allow for more efficient population of higher J CO states and thus, more effective CO cooling. The CO cooling rate is the least temperature sensitive of any of the plotted cooling terms, whereas the H$_2$ cooling rate shows a strong dependence on temperature. The H$_2$ cooling rate is strongly peaked around the temperature peak and shows the most significant change between the two shock models. While the gas-grain cooling rate is temperature dependent, it is also clearly larger at higher densities as the gas-grain cooling curves are skewed toward the higher density sides of the shocks. High frequency noise, which is likely numerical in nature, appears toward the end of both models in the CO cooling rate and thus, a boxcar smoothing algorithm has been applied to the CO cooling rates for distances larger than 0.06 pc in model n30v2b1 and for distances larger than 0.04 pc in model n30v3b1. This noise does not significantly affect our results as it only occurs over very limited spatial scales and occurs only when the cooling rates have decreased significantly, such that the noise does not significantly affect the total cooling rate. Furthermore, this noise only occurs when the temperature has dropped below 10 K, at which point the lack of cosmic-ray heating in our models becomes important (see Section \ref{CO}). The other shock models are qualitatively similar to the ones shown in Figure \ref{fig:profiles} and thus, are not shown. The temperatures, densities, and timescales of the shocks are too low for any significant chemical changes to occur within the gas in any of the models and thus, the small changes in chemical abundance across the shock models are also not shown. 

	The dominant molecular coolant in all of these slow shock models is $^{12}$CO, with 40\%-80\% of the dissipated energy going into $^{12}$CO rotational lines. A significant fraction, 15\%-60\%, of the dissipated energy is not radiated away, but rather, goes toward compressing the magnetic field. In the models with the weakest shocks, those with b = 0.3 and a shock velocity of 2 km s$^{-1}$, the conversion of kinetic energy into magnetic energy is the most significant mechanism for dissipating kinetic energy. Molecular hydrogen rotational lines are the second most effective molecular coolant, but dissipate less than 1\% of the shock energy in all but the models with b = 0.1 and a shock velocity of 3 km s$^{-1}$, which are the models with the strongest shocks. In these stronger shock models, H$_2$ lines account for between 7\% and 21\% of the energy dissipated, with H$_2$ cooling being more important at lower densities. All other cooling mechanisms are very minor in these shock models. A summary of where the energy goes in each model is given in Table \ref{table:ediss}. It should be noted that the sums of the cooling functions do not exactly equal the total kinetic energy dissipation rates. The total cooling rate, however, is never more than 5\% discrepant from the kinetic energy dissipation rate. We believe that this discrepancy arises from difficulties in extending our shock cooling functions to low temperature but do not believe that this small discrepancy significantly affects our results.
		
\begin{deluxetable}{cccccc}
\tabletypesize{\scriptsize}
\tablecolumns{6}
\tablecaption{Sources of Energy Dissipation in the Shock Models \label{table:ediss}}
\tablewidth{0pt}
\tablehead{
\colhead{Model} & \colhead{$E_{CO}$} & \colhead{$E_{B}$} & \colhead{$E_{H_2}$} & \colhead{$E_{dust}$} & \colhead{$E_{H_2O}$} \\
\colhead{(1)} & \colhead{(2)} & \colhead{(3)} & \colhead{(4)} & \colhead{(5)} & \colhead{(6)} 
}
\startdata
n25v2b1 & 76 & 23 & 0.8 & 2 & 0.02 \\
n25v3b1 & 61 & 17 & 21 & 2 & 0.02 \\
n25v2b3 & 42 & 56 & $<$0.1 & $<$0.1 & 0.01\\ 
n25v3b3 & 54 & 44 & 0.9 & 1 & 0.01 \\
n30v2b1 & 76 & 26 & 0.2 & 2 & 0.02 \\
n30v3b1 & 68 & 18 & 14 & 3 & 0.03 \\
n30v2b3 & 43 & 54 & $<$0.1 & 0.2 & 0.01\\
n30v3b3 & 55 & 41 & 0.3 & 2 & 0.02 \\
n35v2b1 & 74 & 25 & 0.1 & 3 & 0.04 \\
n35v3b1 & 72 & 19& 7 & 4 & 0.04 \\
n35v2b3 & 41 & 53 & $<$0.1 & 1 & 0.03\\
n35v3b3 & 53 & 42 & 0.1 & 3 & 0.03 
\enddata
\tablecomments{Column 1 shows the model names. Columns 2-6, respectively, list the percentage of the kinetic energy of the shock that is dissipated via CO rotational lines, increasing the magnetic field strength, H$_2$ lines, gas-grain collisions, and H$_2$O lines.}
\end{deluxetable}
	
	The 12 modeled shocks are relatively weak shocks and produce density enhancements of at most a factor of $\sim$20, of the order of the Mach number. Such compressions are still smaller than the density contrast between the ambient material in molecular clouds and dense cores, which have densities of 10$^5$ cm$^{-3}$ or above (e.g., \citealt{DiFrancesco07}). The maximum temperature in the shocked gas varies significantly, with the stronger shock models achieving maximum temperatures of approximately 150 K and the weaker shock models not even warming up to 20 K. The maximum density and temperature reached in each model is given in Table \ref{table:final}.
	
	The cooling length of each shock is taken to be the full width at quarter maximum of the total cooling function profile. The cooling lengths range from 0.01 pc to 0.35 pc, with the high magnetic field strength and low-density models having the largest cooling lengths. The corresponding cooling timescales range from $5 \times 10^3$ years to $3 \times 10^5$ years, with the longer cooling timescales corresponding to larger cooling lengths. 
	
\begin{figure}[htbp] 
   \centering
   \includegraphics[width=3.5in]{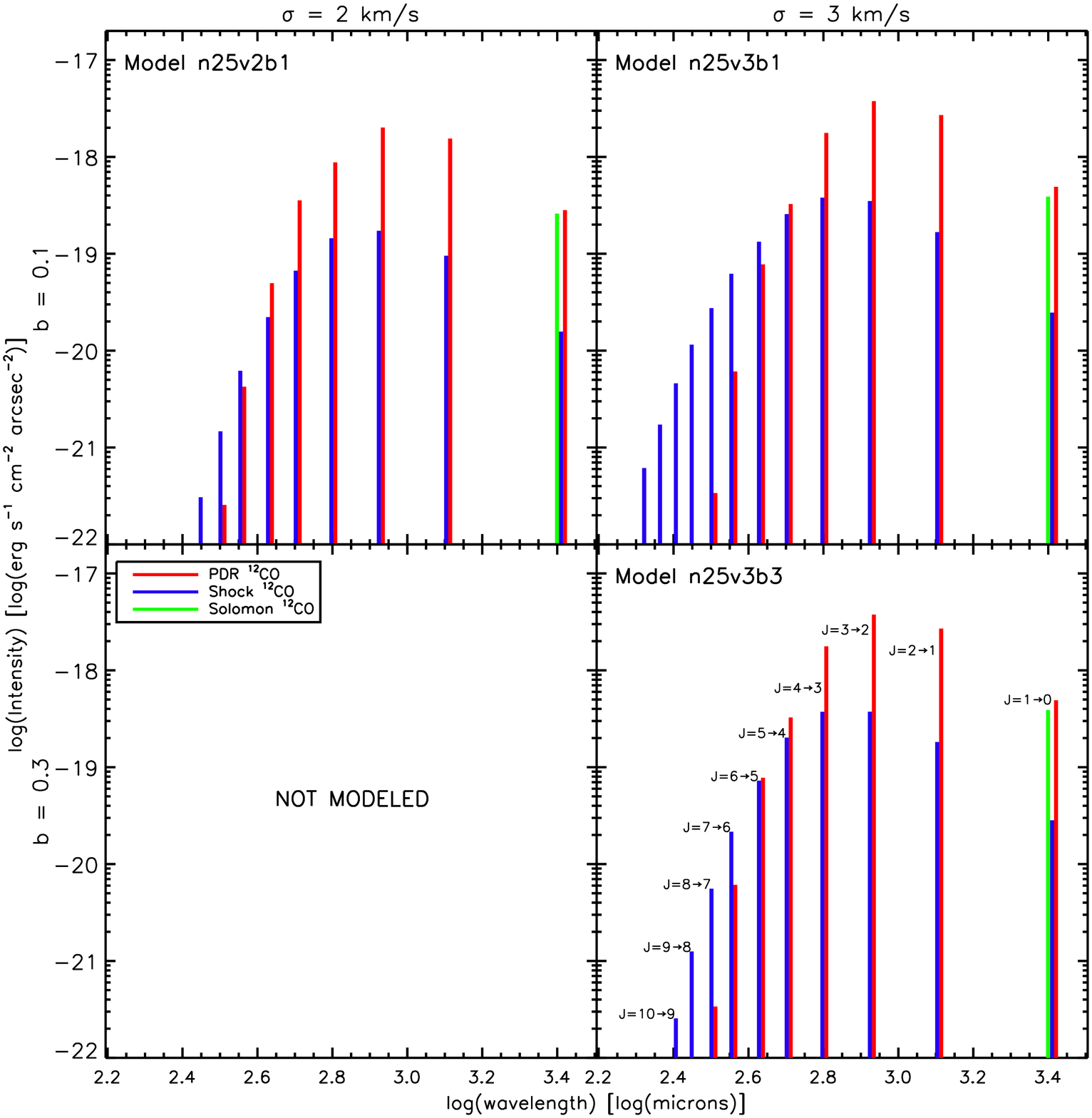}
   \caption{Integrated intensities of various $^{12}$CO rotational transitions for shock models with densities of 10$^{2.5}$ cm$^{-3}$ in units of erg s$^{-1}$ cm$^{-2}$ arcsec$^{-2}$. The shock velocity and magnetic field b parameter used for each shock model are given on the top and left of the grid, respectively, while the model name is given in the top left hand corner of each box. The green (lightest) lines show the \citet{Solomon87} $^{12}$CO $\mbox{J} = 1 \rightarrow 0$ line strengths, the blue (darkest) lines show the CO shock spectra, and the red (medium) lines show the CO PDR spectra. The ten lowest rotational transitions of CO are labeled in the lower right grid panel. Note how the shock spectra dominate over the PDR spectra for high J transitions.}
   \label{fig:specn25}
\end{figure}

\begin{figure}[htbp] 
   \centering
   \includegraphics[width=3.5in]{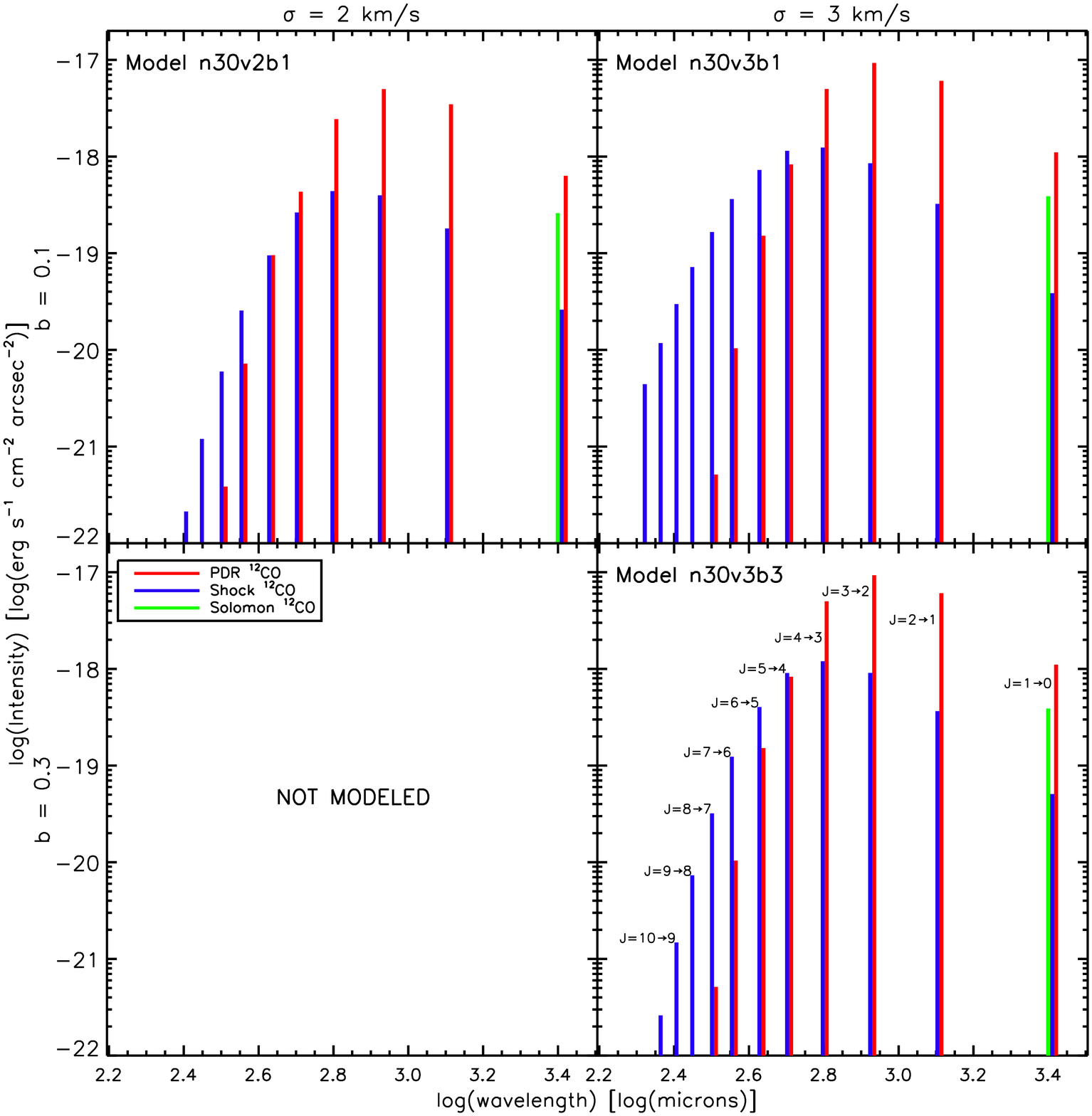}
   \caption{Integrated intensities of various $^{12}$CO rotational transitions for shock models with densities of 10$^{3}$ cm$^{-3}$ in units of erg s$^{-1}$ cm$^{-2}$ arcsec$^{-2}$. The shock velocity and magnetic field b parameter used for each shock model are given on the top and left of the grid, respectively, while the model name is given in the top left hand corner of each box. The green (lightest) lines show the \citet{Solomon87} $^{12}$CO $\mbox{J} = 1 \rightarrow 0$ line strengths, the blue (darkest) lines show the CO shock spectra, and the red (medium) lines show the CO PDR spectra. The ten lowest rotational transitions of CO are labeled in the lower right grid panel. Note how the shock spectra dominate over the PDR spectra for high J transitions.}
   \label{fig:specn3}
\end{figure}

\begin{figure}[htbp] 
   \centering
   \includegraphics[width=3.5in]{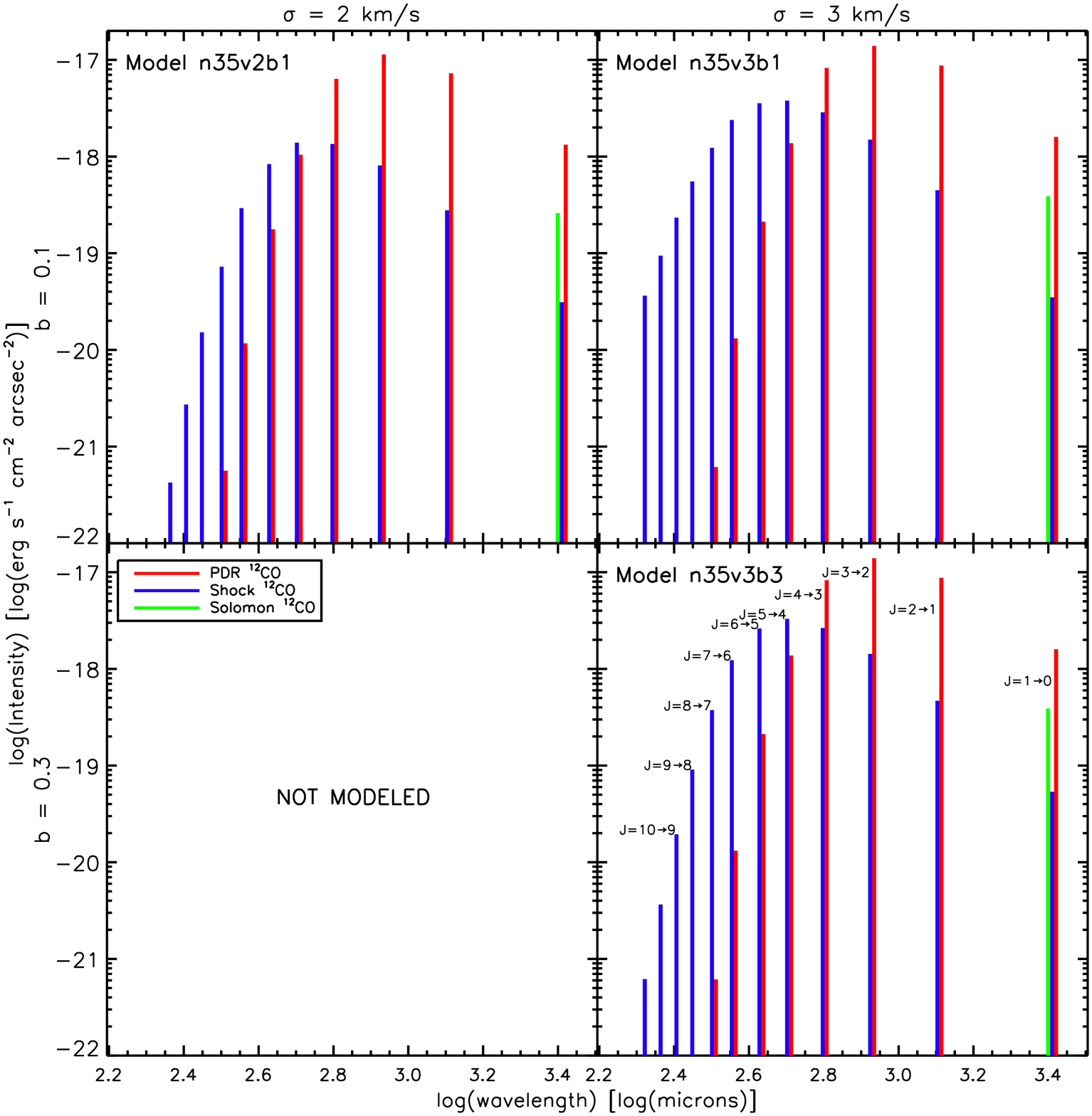}
   \caption{Integrated intensities of various $^{12}$CO rotational transitions for shock models with densities of 10$^{3.5}$ cm$^{-3}$ in units of erg s$^{-1}$ cm$^{-2}$ arcsec$^{-2}$. The shock velocity and magnetic field b parameter used for each shock model are given on the top and left of the grid, respectively, while the model name is given in the top left hand corner of each box. The green (lightest) lines show the \citet{Solomon87} $^{12}$CO $\mbox{J} = 1 \rightarrow 0$ line strengths, the blue (darkest) lines show the CO shock spectra, and the red (medium) lines show the CO PDR spectra. The ten lowest rotational transitions of CO are labeled in the lower right grid panel. Note how the shock spectra dominate over the PDR spectra for high J transitions.}
   \label{fig:specn35}
\end{figure}

	The volume filling factor of shocked gas in a molecular cloud, $ff$, can be calculated from
\begin{equation}
ff = \frac{\Gamma_{turb} \, d_{cool}}{\Delta E_k},
\end{equation}
where $\Gamma_{turb}$ is the turbulent energy dissipation rate per volume (given by Equation (\ref{eqn:gammaturb2})), $d_{cool}$ is the cooling length of the shock, and $\Delta E_k$ is the kinetic energy dissipated per shock front area. Both $d_{cool}$ and $\Delta E_k$ are calculated by the shock code but the cloud radius is required to calculate $\Gamma_{turb}$. For this filling factor calculation, we use the relatively well-established size-velocity relation of Larson's laws, Equation (\ref{eqn:larson1}), to determine the appropriate cloud radius for each shock model. We reiterate, however, that the integrated intensities presented in Figures \ref{fig:specn25}-\ref{fig:specn35} and in Table \ref{table:h2int} are derived independently from any part of Larson's laws. The volume filling factor of shocked gas is always between 0.02\% and 0.5\% of the cloud volume, except for the three weakest shock models, where the volume filling factor becomes as large as 2\%. The cooling times, cooling lengths, and filling factors for all 12 models are given in Table \ref{table:final}. 

\begin{deluxetable}{cccccccc}
\tabletypesize{\scriptsize}
\tablecolumns{8}
\tablecaption{Shock Model Properties \label{table:final}}
\tablewidth{0pt}
\tablehead{
\colhead{Model} & \colhead{log(n$_{max}$)} & \colhead{$T_{max}$} & \colhead{$t_{cool}$} & \colhead{$d_{cool}$} & \colhead{ff} & \colhead{$R_B$} & \colhead{$R_{AD}$}\\
\colhead{} & \colhead{(cm$^{-3}$)} & \colhead{(K)} & \colhead{($10^{4}$ years)} & \colhead{(pc)} & \colhead{(\%)} & \colhead{} & \colhead{}\\
\colhead{(1)} & \colhead{(2)} & \colhead{(3)} & \colhead{(4)} & \colhead{(5)} & \colhead{(6)} & \colhead{(7)} & \colhead{(8)} 
}
\startdata
n25v2b1 & 3.7 & 56 & 6.4 & 0.06 & 0.38 & 14 & 54\\
n25v3b1 & 3.8 & 145 & 3.0 & 0.04 & 0.11 & 22 & 104\\
n25v2b3 & 3.1 & 11 & 28.5 & 0.35 & 2.15 & 4 & 10\\
n25v3b3 & 3.3 & 60 & 9.8 & 0.17 & 0.46 & 7 & 19\\
n30v2b1 & 4.1 & 54 & 2.6 & 0.03 & 0.16 & 14 & 57\\
n30v3b1 & 4.3 & 154 & 1.3 & 0.02 & 0.05 & 22 & 106\\
n30v2b3 & 3.6 & 13 & 11.8 & 0.14 & 0.87 & 4 & 9\\
n30v3b3 & 3.8 & 60 & 3.9 & 0.07 & 0.18 & 7 & 18\\
n35v2b1 & 4.6 & 53 & 1.1 & 0.01 & 0.07 & 14 & 55\\
n35v3b1 & 4.7 & 157 & 0.5 & 0.01 & 0.02 & 21 & 108\\
n35v2b3 & 4.1 & 17 & 4.9 & 0.06 & 0.36 & 4 & 9\\
n35v3b3 & 4.3 & 61 & 1.7 & 0.03 & 0.08 & 7 & 18
\enddata
\tablecomments{Column 1 shows the model names while Columns 2 and 3 show the logarithm of the maximum density reached and the maximum temperature reached, respectively. Column 4 shows the cooling time of the shocked gas and Column 5 shows the cooling length. The volume filling factor of shocked gas for a cloud compatible with the size-velocity relationship of molecular clouds is given in Column 6. Columns 7 and 8 show the factors by which the magnetic field strength and the ambipolar diffusion volume heating rate increase between the initial and shocked gas.}
\end{deluxetable}

Figure \ref{fig:specn25} shows the integrated intensities of CO rotational transitions as calculated from the shock models with densities of 10$^{2.5}$ cm$^{-3}$. The CO spectra from the corresponding comparison PDR models, as well as the estimates for the J = 1 $\rightarrow$ 0 integrated intensities from the \citet{Solomon87} scaling relation, are also shown in Figure \ref{fig:specn25}. Figures \ref{fig:specn3} and \ref{fig:specn35} show the shock spectra from the models with densities of 10$^{3}$ cm$^{-3}$ and 10$^{3.5}$ cm$^{-3}$ respectively, as well as the corresponding PDR spectra and \citet{Solomon87} J = 1 $\rightarrow$ 0 integrated intensities. All integrated intensities are given in units of erg s$^{-1}$ cm$^{-2}$ arcsec$^{-2}$. The y-axis ranges in Figures \ref{fig:specn25}, \ref{fig:specn3}, and \ref{fig:specn35} are identical to facilitate comparisons between the different shock models. As explained in further detail in Section \ref{CO}, CO spectra have not been calculated for any shock models with b = 0.3 and a shock velocity of 2 km s$^{-1}$.

	In all of the cases considered, the PDR emission is significantly stronger, by approximately an order of magnitude, than the shock emission in the three lowest CO rotational transitions. The \citet{Solomon87} $^{12}$CO J = 1 $\rightarrow$ 0 integrated intensity is also larger than the predicted shock J = 1 $\rightarrow$ 0 integrated intensity in all of the models. On the other hand, in all of the models, the shock emission is stronger for all of the high J transitions.  In the strongest, high-density shock models, the shock-integrated intensity becomes larger than the PDR-integrated intensity in the J = 5 $\rightarrow$ 4 line. The PDR emission in the higher J lines drops much more rapidly than the shock emission and the shock emission is an order of magnitude stronger in the J = 7 $\rightarrow$ 6 line in most models. 
	
	While the shock code does not calculate the individual strengths of the different H$_2$ rotational transitions, the lowest rotational levels should be in local thermodynamic equilibrium (LTE) at the modeled densities. Thus, it is assumed that all of the H$_2$ line ratios are given by their LTE values. At 50 K, the ratio of S(1) to S(0) is only $3.3 \times 10^{-3}$, but at 150 K, this ratio increases to 2.3. The S(2) and higher lines are negligible in comparison to the S(1) and S(0) lines at these temperatures. For the three models with significant H$_2$ emission, the expected integrated intensities of the S(1) and S(0) lines are given in Table \ref{table:h2int}.

While the PDR models used do not include the expected line strengths for H$_2$ rotational transitions, \citet{Kaufman06} found that the S(1) and S(0) H$_2$ lines have integrated intensities of $7 \times 10^{-20}$ erg s$^{-1}$ cm$^{-2}$ arcsecond$^{-2}$ and $2 \times 10^{-19}$ erg s$^{-1}$ cm$^{-2}$ arcsecond$^{-2}$, respectively, in a PDR with a density of $10^3$ cm$^{-3}$ and an ISRF of 3 Habing. Therefore, the PDR H$_2$-integrated intensities are expected to be comparable to or slightly lower than the shock H$_2$-integrated intensities.

\begin{deluxetable}{ccc}
\tablecolumns{3}
\tablecaption{Predicted H$_2$-Integrated Intensities \label{table:h2int}}
\tablewidth{0pt}
\tablehead{
\colhead{Model} & \colhead{S(0)} & \colhead{S(1)} \\
\colhead{} & \colhead{(10$^{-19}$ erg s$^{-1}$} & \colhead{(10$^{-19}$ erg s$^{-1}$}  \\
\colhead{} & \colhead{cm$^{-2}$ arcsec$^{-2}$)} & \colhead{cm$^{-2}$ arcsec$^{-2}$)}  \\
\colhead{(1)} & \colhead{(2)} & \colhead{(3)}
}
\startdata
n25v3b1 & 1.5 & 3.5 \\
n30v3b1 & 3.0 & 6.9 \\
n35v3b1 & 5.0 & 11 
\enddata
\tablecomments{Column 1 shows the model names while Columns 2 and 3 show the integrated intensities of the S(0) and S(1) lines, respectively, for the three shock models with significant H$_2$ cooling.}
\end{deluxetable}

\section{DISCUSSION}
\label{discussion}

\subsection{CO Lines}
\label{CO}

	As mentioned in Section \ref{model emission}, no CO spectra are calculated for the models with b = 0.3 and a shock velocity of 2 km s$^{-1}$. This is because the maximum post shock temperatures in these three models, 11, 13, and 17 K, are on the order of the temperature expected from cosmic-ray heating alone (e.g., \citealt{Goldsmith01, Pan09}). The shock code used does not contain any additional heating sources, such as cosmic-ray or photoelectric effect heating, and thus, underestimates the temperatures that would be achieved in these weak shocks. The combination of shock heating and cosmic-ray heating must produce temperatures higher than that produced by cosmic-ray heating alone. Note that in Figure \ref{fig:profiles}, the gas temperature is initially less than 0.1 K and falls below 10 K by the end of the model due to the lack of these extra heating terms. As such, we do not consider the models with b = 0.3 and a shock velocity of 2 km s$^{-1}$ to be valid. The lack of cosmic-ray heating does not significantly affect the other models because the peak temperatures are much larger than 10 K in these models, indicating that shock heating is significantly stronger than what cosmic-ray heating would be. 

	Figures \ref{fig:specn25}-\ref{fig:specn35} show that almost all of the emission from a molecular cloud in the lowest three rotational transitions of CO comes from unshocked gas. These figures also show, however, that most of the emission coming from the mid to high J CO lines (lines at or above J = 7 $\rightarrow$ 6) comes from shocked gas. Not only should the integrated intensities of these lines be higher than predicted from PDR models, but the excitation temperatures derived from the line ratios of these lines should also be higher than a PDR model would predict. Thus, these mid to high J CO lines should serve as observational diagnostics of turbulent energy dissipating via shocks. Since CO accounts for the majority of the cooling via shocks, these mid to high J transitions of CO should be the best tracers of where the majority of the energy goes in turbulent shocks in molecular clouds. 

	Shock emission dominates at higher J transitions because the CO in the shocked gas is warmer than the majority of the CO in the rest of the cloud. While the outer layers of a molecular cloud can be quite warm, due to the incident ISRF, there is little CO in these warm outer regions. The majority of the CO flux in the PDR models comes from gas that is below 20 K. Since our low-velocity shocks are C shocks, CO survives the shock and radiates from gas at temperatures in excess of 50 K. 

	The \citet{Solomon87} $^{12}$CO J = 1 $\rightarrow$ 0 integrated intensity is larger than the predicted shock-integrated intensity in all of the models, which confirms that shock emission is not the major source of emission in the 1 $\rightarrow$ 0 transition. While the comparison PDR models for the high-density shock models over predict the 1 $\rightarrow$ 0 integrated intensity, compared to the \citet{Solomon87}-integrated intensity, the comparison PDR models for the 10$^{2.5}$ cm$^{-3}$ shock models have J = 1 $\rightarrow$ 0 integrated intensities in remarkably good agreement with the \citet{Solomon87}-integrated intensity. This slight discrepancy between some of the PDR models and the \citet{Solomon87} relation is likely due to the characteristic line width, depth, and density of the clouds observed by \citet{Solomon87} being slightly different than the values used in the PDR models.

\subsection{Variation across Parameter Space}
\label{parameter space}
	
	The maximum temperature reached in the shocked gas increases with increasing Alfv\'{e}nic Mach number and Mach number of the shock. All of the models with a shock velocity of 3 km s$^{-1}$, and thus a Mach number of 17, have higher maximum temperatures than all of the models with a shock velocity of 2 km s$^{-1}$, corresponding to a Mach number of 12. For models with the same Mach number, models with higher Alfv\'{e}nic Mach numbers, those with lower magnetic b parameters, have higher maximum temperatures. A larger Alfv\'{e}nic Mach number alone, however, does not necessarily imply a larger maximum temperature, as the models with b = 1 and a shock velocity of 2 km s$^{-1}$ have roughly the same maximum temperature as the models with b = 3 and a shock velocity of 3 km s$^{-1}$ despite having almost twice as large Alfv\'{e}nic Mach numbers. The maximum temperature reached significantly affects the CO shock profile, as higher temperatures excite higher rotational states, which leads to considerably more emission in higher J transitions. As described further in Section \ref{h2}, the effectiveness of H$_2$ cooling is also highly sensitive to the maximum temperature reached in the gas. 
	
	The fraction of energy going toward compressing the magnetic field is primarily dependent upon the Alfv\'{e}nic Mach number of the shock, with more energy going into the magnetic field in the models with lower Alfv\'{e}nic Mach numbers. Larger Alfv\'{e}nic Mach numbers also produce smaller cooling lengths, cooling times, and filling factors. 
	
	The turbulent energy density of a molecular cloud is dependent upon both the density and turbulent velocity dispersion of the cloud. Thus, the models with higher densities and higher shock velocities have larger integrated intensity scaling factors (see Equation (\ref{eqn:intturb})) and, consequently, higher integrated intensities for all CO transitions.

	The critical densities for all the CO lines above the J = 3 $\rightarrow$ 2 line are greater than 10$^5$ cm$^{-3}$ \citep{Schoier05}. Therefore, larger initial densities make it easier for higher rotational states to be populated and, as such, more emission comes out in higher lying lines in the higher density models. This effect is clearly seen in the shock models with b = 0.1 and a shock velocity of 2 km s$^{-1}$, as the line with the largest integrated intensity shifts from the 3 $\rightarrow$ 2 line to the 4 $\rightarrow$ 3 and then finally to the 5 $\rightarrow$ 4 line as the density increases from 10$^{2.5}$ cm$^{-3}$ to 10$^{3}$ cm$^{-3}$ and then to 10$^{3.5}$ cm$^{-3}$. The peak of the PDR spectra remains at the 3 $\rightarrow$ 2 transition in all of the PDR comparison models because the same density is used for all of the PDR comparison models. This difference between the shock and PDR models causes the transition at which shock emission becomes larger than PDR emission to move to slightly lower transitions as the density of the shock model increases. 

As described above, changes to the magnetic field strength, shock velocity, or initial gas temperature can significantly alter the shape and the scaling of the shock CO spectrum. None of these changes, however, alter the key result that shock emission dominates at mid to high J transitions, particularly from J = 7 $\rightarrow$ 6 and up.

The fraction of energy dissipated via CO cooling is not strongly correlated with any shock property. Only in the strongest shock models does the fraction of energy dissipated via CO weakly depend upon the initial density of the gas. For these strong shock models, the fraction of energy that is emitted via CO rotation lines increases by approximately 10\% as the initial density increases from 10$^{2.5}$ cm$^{-3}$ to 10$^{3.5}$ cm$^{-3}$. 

An increase in the initial density of the gas weakly increases the effectiveness of gas-grain cooling, but this cooling term accounts for at most 4\% of the shock cooling in any of the models. Much higher densities, densities closer to 10$^5$ cm$^{-3}$, are required before gas-grain coupling becomes reasonably efficient. Going from a density of 10$^{2.5}$ cm$^{-3}$ to 10$^{3.5}$ cm$^{-3}$ also decreases the cooling lengths, cooling times, and filling factors of all of the models by approximately a factor of six.

\subsection{Magnetic Field Compression}
\label{B field}

In our low velocity shock models, the energy that goes into compressing the magnetic field is of the same order of magnitude as the energy radiated away in CO rotational lines. It is, however, unclear where this injected magnetic energy would go. A local increase in magnetic field strength could further drive MHD waves, which would subsequently shock. In this case, more shocks would be required in order for all of the cloud's turbulent energy to be dissipated by CO cooling and our predicted line integrated intensities would have to be increased by a factor of two. 

Alternatively, this magnetic energy may slowly leak out of the cloud via magnetic coupling with the external medium, as described by \citet{Elmegreen85} and seen in simulations by \citet{Eng02}. This magnetic energy may also be dissipated on small scales via a process such as ambipolar diffusion. 

\subsection{H$_2$ Lines}
\label{h2}

Molecular hydrogen does not have a permanent dipole moment and thus, radiates through weak quadrupole transitions ($\Delta$J = 2). Furthermore, since hydrogen is so light, the rotational energy levels of molecular hydrogen are relatively widely spaced. These two effects make H$_2$ rotational emission highly temperature sensitive and temperatures in excess of 100 K are required for significant emission. This temperature sensitivity can be seen in the shock models, as H$_2$ emission is essentially negligible in all but the strongest shock models, which are the only models where the maximum temperature exceeds 100 K. In these models, the S(1) and S(0) H$_2$ lines have comparable integrated intensities to the CO J = 7 $\rightarrow$ 6 line, although the H$_2$ lines are relatively stronger in the lower density models.

The lack of H$_2$ rotational emission from cool gas means that, aside from shocked gas, the only significant source of H$_2$ emission in a molecular cloud is the thin outer edge of the cloud's PDR where the temperature is high and H$_2$ is not rapidly photodissociated. Thus, H$_2$ rotational emission could be a very useful tracer of shocked gas in a molecular cloud. In particular, while the predicted S(0) shock-integrated intensities are on the order of the S(0) PDR-integrated intensity, the S(1) shock-integrated intensities range from being five times to over fifteen times larger than the S(1) PDR-integrated intensity. The ratio of S(1)/S(0) at the 150 K maximum temperature of the strongest shocks, approximately 2.3, is also significantly different from the ratio of these lines in the \citet{Kaufman06} comparison PDR models, roughly 0.3. 

H$_2$ cooling may be even more significant in gas that is lacking in gas phase CO, as this shocked gas is likely to reach higher temperatures with the effectiveness of CO cooling reduced. This increase in H$_2$ shock emission in CO sparse gas, such as the ``dark gas'' in the periphery of a molecular cloud (e.g., \citealt{Wolfire10}), may naturally produce a limb brightening effect for H$_2$ rotational emission in molecular clouds, as observed in Taurus by \citet{Goldsmith10}.

In all of our shock models, cooling from vibrational transitions of H$_2$ is negligible because temperatures on the order of a few thousand Kelvin are required to excite higher energy vibrational states \citep{Kaufman96II}.

\subsection{Other Shock Tracers}
\label{shock tracers}

Water is known to be an effective coolant in high-velocity shocks \citep{Kaufman96II} but water cooling is negligible in all of our shock models. This is because water is only formed in gas with a temperature of a few hundred Kelvin \citep{Elitzur78deJong,Elitzur78Watson} and is only efficiently liberated from dust grains, due to sputtering, in 15 km s$^{-1}$ or faster shocks \citep{Draine83}. Low-velocity shocks are also very ineffective at heating dust grains, meaning that thermal sublimation of water off of dust grains is completely negligible in our low-velocity shocks \citep{Draine83}. 

At the low densities of our shock models, the interaction timescale of gas with dust grains is long compared to the cooling time such that only a few percent of the energy is ever liberated via gas-grain interactions in the models. Densities closer to 10$^5$ cm$^{-3}$ are needed before gas-grain coupling becomes effective.

While CO line radiation and the compression of magnetic fields are the dominant coolants in our low-velocity shocks, other molecular lines, which have not been included in our shock models, may still be valuable tracers of shocked gas. In particular, molecular lines that are sensitive to increased temperature could provide shock tracers in different wavelength regimes. Molecular transitions that are sensitive to density may also be useful shock tracers, but are less likely to be as useful as temperature sensitive lines since the maximum densities reached in the shocked gas are less than the gas density of prestellar cores.

\subsection{Additional Caveats}
\label{caveats}
		
	In scaling the shock models to predict the total strength of the shock emission from an entire molecular cloud, it was assumed that all of the turbulent energy of the cloud is dissipated at one particular shock strength. In reality, energy will be dissipated through a variety of different strength shocks, either due to different shock velocities or different strengths of the magnetic field parallel to the shock front. Furthermore, \citet{SmithHeitsch00} show that high-velocity shocks dominate energy dissipation in driven turbulence while \citet{SmithZuev00} show that if the turbulence is decaying, low-velocity shocks dissipate most of the energy. If energy is dissipated through lower velocity shocks, then our calculated line-integrated intensities will overestimate the actual emission from molecular clouds in lines with higher excitation temperatures. If turbulence is driven at scales much smaller than the size of the cloud, however, then our line strengths should be increased by a factor of $\kappa^{-1}$ (see Section \ref{lturb} for a discussion on how $\kappa$ relates to the driving scale of turbulence).
	
	Another factor of two uncertainty in the shock-integrated intensities comes from the assumption that the energy in magnetic field fluctuations in the cloud is negligible. While some MHD simulations have shown that the turbulent kinetic energy dominates over the energy in magnetic field fluctuations, particularly for smaller initial magnetic fields \citep{Padoan99, Padoan00,Heitsch01}, other simulations find that these magnetic field fluctuations have energy on the order of the kinetic energy of the turbulence \citep{Stone98, Ostriker01}. If the energy in magnetic field fluctuations is on a par with the kinetic energy of turbulence, the line-integrated intensities would have to be scaled up by a factor of two to account for the dissipation of this additional energy. 

	In scaling up the shock models to estimate the total integrated intensities from an entire cloud, the integrated intensity of every line in a particular model has been multiplied by the same factor and no optical depth effects have been taken into account. These optical depth effects, however, should not be significant for the higher rotational transitions of CO, where the CO transitions are effective at tracing shock emission, as the PDR models indicate that the CO lines are only optically thick up to, and including, the J = 5 $\rightarrow$ 4 transition. As for shock emission in the lower rotational transitions of CO, this emission is likely to be absorbed by the ambient gas and thereby will serve as a heating source for the ambient gas. Note, however, that the expected CO-integrated intensity at these low transitions is much less than the PDR-integrated intensity, which indicates that this extra shock heating will have only a very minor effect on the temperature, and thus the spectrum, of the ambient gas. 
	
	In deriving the total turbulent energy dissipation rate of a molecular cloud, it was assumed that turbulence decays in a crossing time. Recently, \citet{Basu09} and \citet{Basu10} discovered a long-lived magnetic-tension-driven mode in their thin disk simulations of flattened molecular clouds, arising from interactions between the disk and an external magnetic field, which was able to preserve a significant fraction of the turbulent energy of the cloud for much longer than the crossing time. The existence of such a long-lived MHD mode would significantly reduce the required energy dissipation rate of a molecular cloud and, therefore, our predicted line-integrated intensities as well. This long-lived mode, however, has not been noticed in any further simulations, including the three-dimensional simulations of collapsing cores done by \citet{Kudoh11}, who did look for this particular MHD mode. \citet{Kudoh11} suggest that the absence of this mode in their simulations may be due to the very small density contrast between the disks and surrounding gas in their simulations.
	
	The line ratios from these shock models are independent of the scaling of the lines and thus, are not affected by the above uncertainties. The line ratios of the shock models are also independent of the assumption of spherical geometry.

	The turbulent energy of molecular clouds may also not be dissipated completely through shocks. \citet{SmithZuev00} and \citet{Stone98} find, in their simulations of turbulence, that only 50\% of the turbulent energy is dissipated through artificial viscosity, due to the presence of shock fronts, while the other 50\% is dissipated through numerical viscosity, representative of small-scale dissipation distributed relatively uniformly across the cloud. It is possible that some of this uniformly dissipated energy should have been dissipated in weak shocks or vortices that were not resolved in these simulations and thus, we consider that 50\% is only a lower limit for the fraction of energy dissipated in intermittent structures (i.e., shocks). 
		
	An alternative model for turbulent dissipation, where energy is dissipated through magnetized vortices, has also been put forward \citep{Godard09}. In the \citet{Godard09} turbulent dissipation region (TDR) model, small vortices on the order of a few tens of AU heat gas to temperatures of nearly 1000 K via ion-neutral friction. The spectral signature of such a TDR model should be significantly different from our low-velocity shock models because the TDR model produces temperatures much higher than what can be obtained from slow shocks. 

	All of the shock models have been run with standard, roughly solar, metal abundances. In lower metallicity clouds, the abundance of CO will be reduced. The gas phase CO abundance will also be lower at higher densities, due to freeze out of CO onto dust grains (e.g., \citealt{Goldsmith01}), and in the outer layers of molecular clouds, where CO is readily photodissociated (e.g., \citealt{Wolfire10}). The reduction of gas phase CO may lead to higher post shock temperatures, which would change the resulting CO spectrum and could affect which cooling mechanism is dominant. H$_2$ line cooling and the deposition of energy into magnetic fields may also be more important for dissipating kinetic energy in CO poor gas. Further low-metallicity shock models, however, are needed to confirm this. 

	The CO spectrum predicted from the PDR model is dependent upon the chemistry put into the models. In particular, any chemical effects which would increase the temperature in the outer CO layers, such as an increase in polycyclic aromatic hydrocarbon heating, would shift the PDR spectrum toward higher rotational transitions. The PDR models used also have a relatively low ISRF of 3 G$_0$. In active, high-mass star-forming regions, the ambient ISRF is likely to be much higher, due to the presence of previously formed, massive, young stars and thus, the PDR emission from these regions is likely to be much stronger and peaked toward much higher J transitions. As such, our model comparisons should only be used for relatively quiescent, low-mass star-forming regions in which the ambient ISRF is relatively low, rather than in strong UV environments such as the Orion Bar. 
	
\subsection{Observational Potential} 
\label{observational limits}

The total energy dissipated by shocks cannot be directly observed because some of the turbulent energy is not radiated away, but rather goes into increasing magnetic field strengths. Furthermore, many of the low lying CO lines are not readily observable, because these transitions are dominated by emission from unshocked gas. This ambient gas emission, however, does drop rapidly at higher J numbers and the CO J = 6 $\rightarrow$ 5 (691.47308 GHz) and 7 $\rightarrow$ 6 (806.651806 GHz) transitions are dominated by shock emission in most of our models. Thus, these higher rotational CO transitions act as shock tracers and by fitting shock models to the observed strengths of multiple high J CO transitions, the total shock luminosity of a cloud can be estimated.

The Herschel Space Observatory's Heterodyne Instrument for the Far Infrared (HIFI)  and Spectral and Photometric Imaging Receiver (SPIRE) both have the necessary wavelength coverage and sensitivity to be able to detect the CO J = 5 $\rightarrow$ 4, 6 $\rightarrow$ 5, and 7 $\rightarrow$ 6 lines, if they are as bright as we predict in our shock models. These two Herschel instruments also cover the J = 8 $\rightarrow$ 7 wavelength but our predicted line strengths for this transition are at the limits of what could be detected within a few hours of observing time. From the ground, the James Clerk Maxwell Telescope's (JCMT) receiver W is capable of observing at the wavelength of the CO J = 6 $\rightarrow$ 5 transition while the Atacama Pathfinder Experiment (APEX) and the Caltech Submillimeter Observatory (CSO) have instruments that operate in the appropriate wavelength regimes to detect both the 6 $\rightarrow$ 5 and 7 $\rightarrow$ 6 lines. The CO J = 6 $\rightarrow$ 5 and 7 $\rightarrow$ 6 lines also lie within bands 9 and 10, respectively, of the Atacama Large Millimeter Array (ALMA). ALMA, with its superb resolution, may resolve individual shock fronts and thus, provide information regarding the properties of individual shocks.  

Some care must be taken when choosing a location in a molecular cloud to observe, as there are other sources of high J CO line emission that have not included in the comparison PDR models. High-velocity protostellar outflows will generate large shocks that can easily produce temperatures of hundreds of Kelvin. Such strongly shocked gas will radiate significantly in high J lines (e.g., \citealt{Kaufman96I, Kaufman96II}). Embedded high-mass stars will also significantly heat nearby gas and lead to high J line emission. Thus, while turbulent dissipation should occur in active, high-mass star-forming regions, the spectral signatures of low-velocity, turbulence-induced shocks may only be readily detectable in quiescent regions of low-mass star-forming molecular complexes. Water emission can be used as a discriminant between low-velocity, turbulent shocks and the stronger shocks produced by outflows, as little water emission is predicted from our models while stronger shocks are expected to cool significantly via water lines (e.g., \citealt{Kaufman96II}).
 
The temperature sensitivity of the S(0) and S(1) lines of H$_2$, at 28.2 $\mu$m and 17.0 $\mu$m, respectively, makes these lines potential shock tracers. The atmospheric transmission at 28.2 $\mu$m is, however, very poor, and thus, it is extremely difficult to observe the S(0) line with ground-based facilities. The expected weakness of these two lines makes observations of H$_2$ emission even more problematic.
  
\section{GLOBAL HEATING}
\label{heating}

\subsection{Cosmic Rays}
\label{cosmic}
In the cold, well-shielded, central regions of molecular clouds, cosmic-ray heating is believed to be the dominant heating term (e.g., \citealt{Goldsmith01}), but the general cosmic-ray ionization rate in the galaxy is not particularly well known. Estimates for the cosmic-ray ionization rate in dense gas vary from $10^{-16}$ s$^{-1}$ \citep{Caselli98, Liszt03, Doty04} to $10^{-18}$ s$^{-1}$ \citep{Caselli98, Caselli02, Flower07, Hezareh08} but commonly lie between $1 \times 10^{-17}$ s$^{-1}$ and $5 \times 10^{-17}$ s$^{-1}$ \citep{Williams98, Vandertak00, Doty02, Wakelam05, Bergin06, Maret07, Goicoechea09}. While the spread in cosmic-ray ionization rates may simply be due to spatially varying rates (e.g., \citealt{Vandertak06}), some of the scatter is likely due to uncertainties in the chemical models used to determine the ionization rates, as changes in known reaction rates (e.g., \citealt{Dalgarno06}) and the inclusion of non-equilibrium chemistry \citep{Lintott06} have changed the results of ionization rate calculations. 

For this paper, a cosmic-ray ionization rate range of $1 \times 10^{-17}$ s$^{-1}$ to $5 \times 10^{-17}$ s$^{-1}$ is adopted and, since each cosmic-ray ionization deposits approximately 20 eV into the gas \citep{Goldsmith01}, the volume heating rate by cosmic-rays is taken to be 
\begin{equation}
\Gamma_{cr}= 0.3\mbox{ to }1.6 \times 10^{-27} \left(\frac{n}{\mbox{cm}^{-3}}\right) \mbox{erg s}^{-1} \mbox{ cm}^{-3}.
\end{equation}
This cosmic-ray heating rate is shown in Figure \ref{fig:heating}.

\subsection{Turbulent Heating}
\label{turbulent heating}

If the turbulent energy of a cloud is not dissipated through localized shocks, but rather is dissipated relatively uniformly across the cloud, then this turbulent energy dissipation will act more as a general heating mechanism for the cloud and will yield the heating rate given by Equation (\ref{eqn:gammaturb2}). For the sole purpose of comparing the turbulent energy dissipation rate to the cosmic-ray heating rate, both the size-velocity and density-size relationships from \citet{Solomon87} are used to rewrite the turbulent energy dissipation rate in terms of only $\kappa$ and the gas density. Using these two scaling relations, the turbulent energy heating rate can be written as
\begin{equation}
\Gamma_{turb} = 5.86 \times 10^{-25} \, \kappa^{-1} \, \epsilon \left(\frac{n}{10^3 \, \mbox{cm}^{-3}}\right)^{0.5}  \mbox{ erg cm}^{-3},
\end{equation}
where $\epsilon$ is the fraction of the dissipated turbulent energy that acts as a global heating mechanism. The simulations of \citet{SmithZuev00} and \citet{Stone98} both suggest that roughly half of the turbulent energy of a cloud may be dissipated uniformly across a cloud instead of in localized shocks and thus, suggest that $\epsilon = 0.5$ (see Section \ref{caveats} for further discussion on $\epsilon$). Furthermore, the magnetic energy injected by shocks may also decay and lead to a global heating of the cloud. Figure \ref{fig:heating} shows the turbulent heating rates, as a function of density, for $\epsilon = 1$ and $\epsilon = 0.5$. As before, a fixed $\kappa$ value of 1 is used.

\subsection{Ambipolar Diffusion}
\label{ambipolar}

\begin{figure}[htbp] 
   \centering
   \includegraphics[width=3.5in]{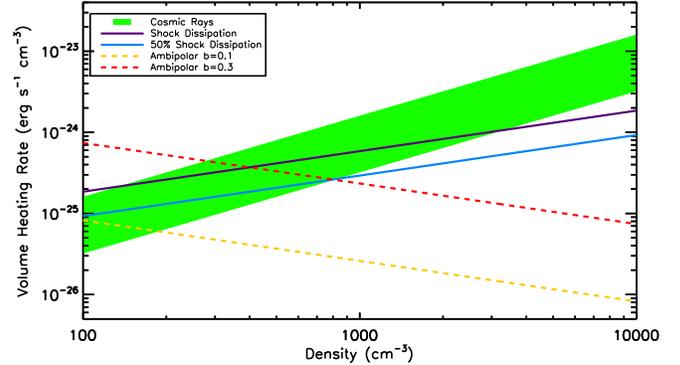}
   \caption{Various heating rates for well-shielded molecular gas. The shaded (green) region shows the range of cosmic-ray heating rates, the dark (purple) solid line shows the total shock turbulent energy dissipation rate, the light (blue) solid line shows 50\% of the shock turbulent dissipation rate, and the dark (red) and light (yellow) dotted lines show the ambipolar diffusion heating rates for b values of 0.3 and 0.1, respectively. Please see the online version for a color version of this figure.}
   \label{fig:heating}
\end{figure}

One potentially important energy dissipation mechanism that is usually not included in simulations (e.g., \citealt{Stone98, MacLow99}) is ambipolar diffusion. From their numerical simulations, \citet{Padoan00} find that the ambipolar heating rate in well shielded molecular gas is given by
\begin{eqnarray}
\Gamma_{ambi} &=& \left(\frac{<|B|>}{10 \, \mu G}\right)^4 \left(\frac{M_A}{5}\right)^2 \left(\frac{<n>}{320 \, \mbox{cm}^{-3}}\right)^{-\frac{3}{2}}  \nonumber \\
                              && 10^{-24} \mbox{ erg cm}^{-3} \mbox{ s}^{-1},
\end{eqnarray}
where $<|B|>$ is the volume-averaged magnetic field strength, $M_A$ is the Alfv\'{e}nic Mach number of the turbulence, and $<n>$ is the volume-averaged number density. Applying Larson's laws and a density-magnetic field strength scaling relation with k = 0.5 converts this equation to the form
\begin{equation}
\Gamma_{ambi} = 2.6 \times 10^{-24} \, b^2 \left(\frac{n}{10^3 \, \mbox{cm}^{-3}}\right)^{-0.5} \mbox{ erg cm}^{-3} \mbox{ s}^{-1}.
\end{equation}
Figure \ref{fig:heating} shows this ambipolar diffusion heating rate, as a function of density, evaluated for the two b parameters previously used for our shock models, b =  0.1 and 0.3. 

\subsection{Discussion}
\label{heating discussion}

	For gas with a density larger than 10$^3$ cm$^{-3}$, the cosmic-ray heating rate is the dominant heating term. The turbulent energy dissipation rate is comparable to the cosmic-ray heating rate around a density of 10$^3$ cm$^{-3}$ and becomes larger than the cosmic-ray heating rate at lower densities. 
	
	\citet{Goldsmith01} examines the thermal balance of molecular clouds and finds that the ambient temperature of well-shielded gas, at a density of 10$^3$ cm$^{-3}$, is 10 K. We took the \citet{Goldsmith01} prescriptions for cooling and heating and increased the cosmic-ray heating by a factor of two to reproduce the effect of having a turbulent energy dissipation heating rate equivalent to the cosmic-ray heating rate. With this increased heating rate, we find a slightly higher equilibrium temperature of 13 K. \citet{Pan09} present a more detailed thermal balance model that also includes a turbulent heating term and, for a 1 pc sized cloud, find similar gas temperatures of 13-17 K, depending upon the cosmic-ray heating rate used.
		
	This small change in temperature is well within the intrinsic scatter of observed gas temperatures in molecular clouds (e.g., \citealt{Bergin07}). Since the cosmic-ray heating rate is also uncertain by at least a factor of two, the finding of previous thermal balance studies that the ambient temperatures of molecular clouds are roughly consistent with heating by cosmic-ray ionization alone (e.g., \citealt{Bergin06}) is not in conflict with the presence of heating from turbulent dissipation at the rate calculated above. The above agreement between observations and models does, however, constrain the turbulent heating rate to not be significantly greater than estimated above. This implies that $\kappa$ cannot be much less than one, similar to the findings of \citet{Basu01}. \citet{Padoan00} also note that if turbulent heating is significant in a molecular cloud, then a positive temperature-velocity dispersion relation is expected, as observed by \citet{Jijina99}. 
		 	
	The significance of ambipolar diffusion to the thermal balance of a cloud is highly dependent upon the strength of the magnetic field. For a strong field, b = 0.3, ambipolar diffusion should be the dominant heating process in well-shielded gas with an average density of 100 cm$^{-3}$, and should be comparable to other heating processes at a density of 10$^3$ cm$^{-3}$. Gas at densities of 100 cm$^{-3}$, however, may not necessarily be well shielded and may instead be dominated by the ISRF. If magnetic field strengths are slightly lower, corresponding to b = 0.1, the ambipolar diffusion rate is negligible for all densities greater than or equal to 100 cm$^{-3}$. It should also be noted that the ambipolar diffusion heating rate becomes larger at lower densities, unlike the turbulent and cosmic-ray heating rates, because the typical collision speed between ions and neutrals increases with decreasing density.
				
	Shocks will cause both the density and magnetic field strength to increase. The change in the density will be proportional to the change in the magnetic field strength, given the flux freezing approximation made in the shock models. The Alfv\'{e}n speed in the gas will thus also increase by a factor proportional to the square root of the change in the magnetic field strength. If the magnetic field strength is locally increased by a factor of q by a shock and the velocity dispersion of a cloud remains relatively unchanged, then the ambipolar diffusion heating rate in the shocked gas will be
\begin{eqnarray}
\Gamma_{ambi,shocked} &=& q^{1.5} \, \Gamma_{ambi,0},
\end{eqnarray}
where $\Gamma_{ambi,0}$ is the ambipolar diffusion dissipation rate in the unshocked gas. Thus, the postshock gas will have its ambipolar diffusion heating rate increased.

Table \ref{table:final} gives the factors by which the magnetic field strength and the ambipolar diffusion heating rate increase in each of the shock models. The magnetic field increases in strength by a factor of 4-22 and the ambipolar diffusion heating rate increases by one to two orders of magnitude. Even for the weakest magnetic field models, the ambipolar diffusion heating rate in the shocked gas is greater than the rate at which energy is injected into the magnetic field through the shock and thus, this enhanced ambipolar diffusion rate may provide an efficient mechanism for dissipating the injected magnetic energy. Furthermore, this enhanced ambipolar diffusion rate should add an extra heating source to the shocked gas, thereby increasing the CO flux that will emerge at higher J transitions.
		 	
The above discussion regarding heating rates is only relevant for the well-shielded centers of molecular clouds, as photoelectric heating due to the ISRF will dominate the heating in the outer PDR zones of molecular clouds (e.g., \citealt{Kaufman99}). The heating rates shown in Figure \ref{fig:heating} were also derived using Larson's laws, which, as discussed in Section \ref{scaling}, have recently been called into question.
	 
\section{CONCLUSIONS}
\label{conclusions}
	We have run models of MHD, C-type shocks, based on \citet{Kaufman96I}, for shock velocities of 2 and 3 km s$^{-1}$ and initial densities between 10$^{2.5}$ and 10$^{3.5}$ cm$^{-3}$. CO is found to be the dominant molecular coolant with 40\%-80\% of the shock energy being emitted in CO rotational lines. All other line cooling processes are negligible, except for H$_2$ line cooling in the models with the very strongest shocks, in which H$_2$ cooling accounts for 5\%-20\% of the total shock cooling. Between 20\% and 60\% of the shock energy also goes into compressing the magnetic field. 
		
The expected CO spectrum from each of the shock models has been calculated and PDR models, based on \citet{Kaufman99}, were used to determine the expected contribution of CO emission from not only the cold, well-shielded interior of a molecular cloud, but also from the warm outer layers of the cloud. The PDR emission dominates for low J transitions of CO but the shock emission is larger at mid to high J transitions. In all models the shock emission is larger than the PDR emission in the J = $7 \rightarrow 6$ transition and the shock emission can dominate as low as the J = 5 $\rightarrow$ 4 transition, depending upon the shock model. The J = 6 $\rightarrow$ 5 and 7 $\rightarrow$ 6 should serve as shock tracers and should be detectable with current observational facilities.
	
	The turbulent energy dissipation rate is larger than the cosmic-ray heating rate for densities less than 10$^3$ cm$^{-3}$. The presence of such an additional heating term is, however, still consistent with previous thermal balance studies, given the uncertainty in the cosmic-ray heating rate and the range of observed molecular cloud temperatures. The ambipolar diffusion heating rate is negligible at high densities and low-magnetic field strengths but can be dominant at lower densities and higher magnetic field strengths. Ambipolar diffusion is also enhanced in the shocked gas and may provide a mechanism for the dissipation of energy injected into the magnetic field by a shock.
			
	We thank Shantanu Basu for helpful suggestions on the role of magnetic fields in molecular clouds as well as our anonymous referee for many useful changes to this paper. A.P. is partially supported by the Natural Sciences and Engineering Research Council of Canada (NSERC) graduate scholarship program and D.J. is supported by a NSERC Discovery Grant. This research has made use of NASA's Astrophysics Data System.
	
\bibliographystyle{apj}

\end{document}